\documentclass[fleqn,usenatbib]{mnras}

\usepackage{newtxtext,newtxmath}

\usepackage[T1]{fontenc}

\DeclareRobustCommand{\VAN}[3]{#2}
\let\VANthebibliography\thebibliography
\def\thebibliography{\DeclareRobustCommand{\VAN}[3]{##3}\VANthebibliography}

\usepackage{graphicx}	
\usepackage{amsmath}	
\usepackage{amssymb}	

\newcommand{\Ni}{\ensuremath{^{56}\mathrm{Ni}}}
\newcommand{\Co}{\ensuremath{^{56}\mathrm{Co}}}
\newcommand{\Fe}{\ensuremath{^{56}\mathrm{Fe}}}
\newcommand{\Msun}{\ensuremath{\mathrm{M}_\odot}}
\newcommand{\Mej}{\ensuremath{M_\mathrm{ej}}}
\newcommand{\Eej}{\ensuremath{E_\mathrm{ej}}}
\newcommand{\Msunpyr}{\ensuremath{\Msun~\mathrm{yr^{-1}}}}
\newcommand{\kmps}{\ensuremath{\mathrm{km~s^{-1}}}}

\graphicspath{{./}{figures/}}

\title[\Ni\ mixing and photospheric velocity]{
Systematic investigation of the effect of \Ni\ mixing in the early photospheric velocity evolution of stripped-envelope supernovae
}

\author[T. J. Moriya et al.]{
Takashi J. Moriya,$^{1,2}$\thanks{E-mail: takashi.moriya@nao.ac.jp (TJM)}
Akihiro Suzuki,$^{1}$
Tomoya Takiwaki,$^{1}$
Yen-Chen Pan,$^{1}$ 
and
\newauthor
Sergei I. Blinnikov$^{3,4,5}$
\\
$^{1}$National Astronomical Observatory of Japan, National Institutes of Natural Sciences, 2-21-1 Osawa, Mitaka, Tokyo 181-8588, Japan \\
$^{2}$School of Physics and Astronomy, Faculty of Science, Monash University, Clayton, Victoria 3800, Australia\\
$^{3}$National Research Center "Kurchatov institute", Institute for Theoretical and Experimental Physics (ITEP), 117218 Moscow, Russia \\
$^{4}$Sternberg Astronomical Institute, M.V. Lomonosov Moscow State University, Universitetski pr. 13, 119234 Moscow, Russia \\
$^{5}$Kavli Institute for the Physics and Mathematics of the Universe (WPI), The University of Tokyo Institutes for Advanced Study, The University of Tokyo, \\
5-1-5 Kashiwanoha, Kashiwa, Chiba 277-8583, Japan
}

\date{Accepted 2020 July 8. Received 2020 July 8; in original form 2020 May 21.}

\pubyear{2020}

\begin{document}
\label{firstpage}
\pagerange{\pageref{firstpage}--\pageref{lastpage}}
\maketitle

\begin{abstract}
Mixing of \Ni, whose nuclear decay energy is a major luminosity source in stripped-envelope supernovae, is known to affect the observational properties of stripped-envelope supernovae such as light-curve and color evolution. Here we systematically investigate the effect of \Ni\ mixing on the photospheric velocity evolution in stripped-envelope supernovae. We show that \Ni\ mixing significantly affects the early photospheric velocity evolution. The photospheric velocity, which is often used to constrain the ejecta mass and explosion energy, significantly varies by just changing the degree of \Ni\ mixing. In addition, the models with a small degree of \Ni\ mixing show a flattening in the early photospheric velocity evolution, while the fully mixed models show a monotonic decrease. The velocity flattening appears in both helium and carbon+oxygen progenitor explosions with a variety of ejecta mass, explosion energy, and \Ni\ mass. 
Some stripped-envelope supernovae with early photospheric velocity information do show such a flattening.
We find that Type~Ib SN~2007Y, which has early photospheric velocity information, has a signature of a moderate degree of \Ni\ mixing in the photospheric velocity evolution and about a half of the ejecta is mixed in it. 
The immediate spectroscopic follow-up observations of stripped-envelope supernovae shortly after the explosion providing the early photospheric evolution give an important clue to constrain \Ni\ mixing in the ejecta.
\end{abstract}




\begin{table*}
	\centering
	\caption{Progenitor properties.}
	\label{tab:progenitors}
	\begin{tabular}{lccccccc} 
		\hline
		name   & ZAMS mass & progenitor mass & progenitor radius    & He core mass & C+O core mass & mass cut & ejecta mass \\
		       & (\Msun)   & (\Msun)         & ($\mathrm{R}_\odot$) & (\Msun)      & (\Msun)       & (\Msun)  & (\Msun) \\		
		\hline
		He3.85 & 14        & 3.85 & 4.78 & 3.85 & 2.04 & 1.35 & 2.5  \\
		CO3.85 & 20        & 3.85 & 0.14 & 0    & 3.85 & 1.35 & 2.5  \\
		He4.90 & 17        & 4.90 & 3.63 & 3.85 & 2.92 & 1.40 & 3.5  \\
		\hline
	\end{tabular}
\end{table*}

\begin{figure}
	\includegraphics[width=\columnwidth]{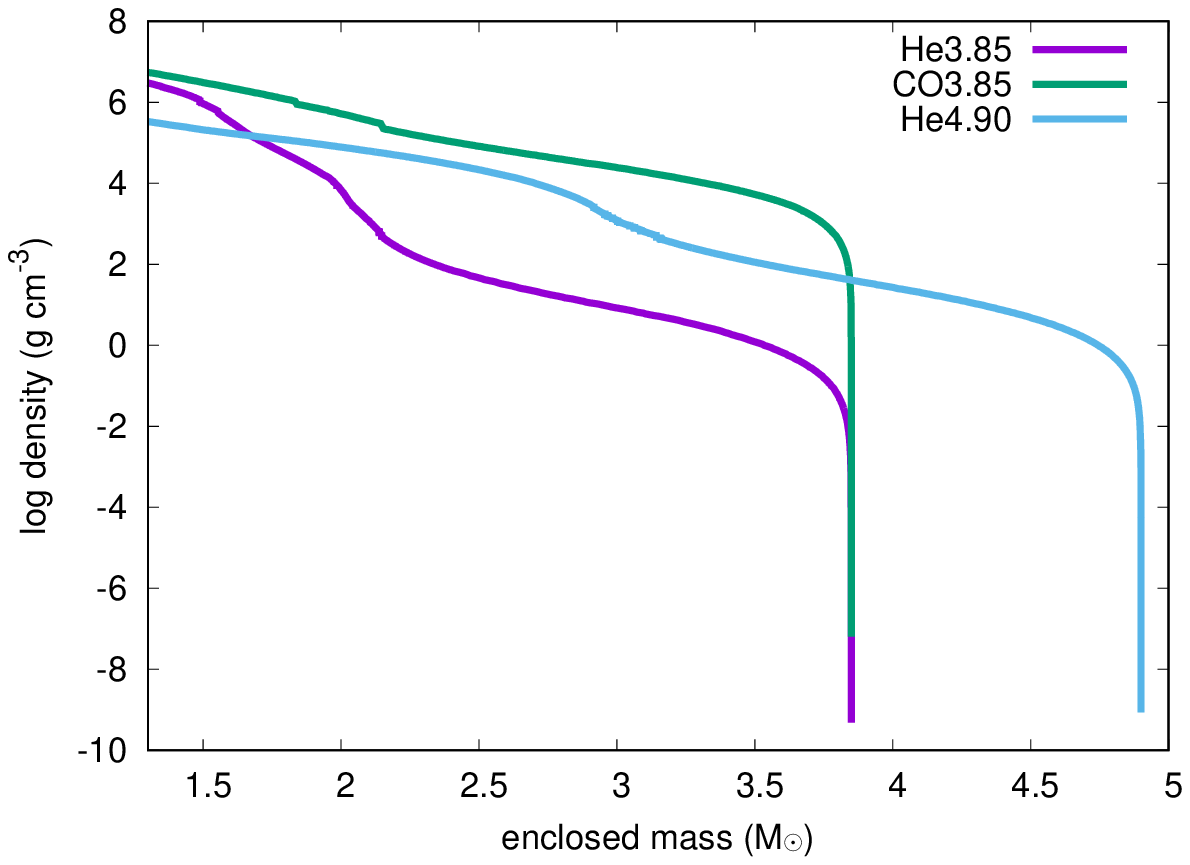}
	\includegraphics[width=\columnwidth]{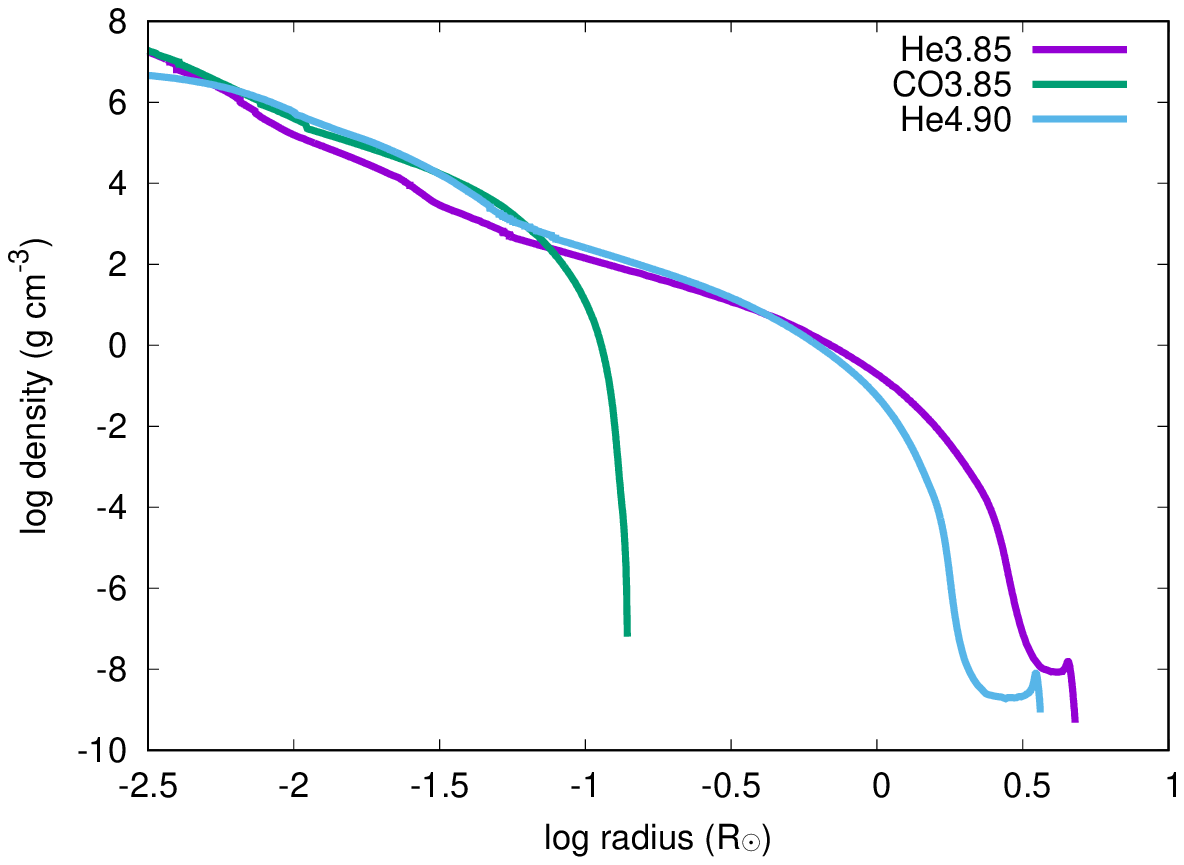}
    \caption{
    Density structure of the progenitor models in the mass coordinate (top) and radius coordinate (bottom).
    }
    \label{fig:density}
\end{figure}

\begin{figure}
	\includegraphics[width=\columnwidth]{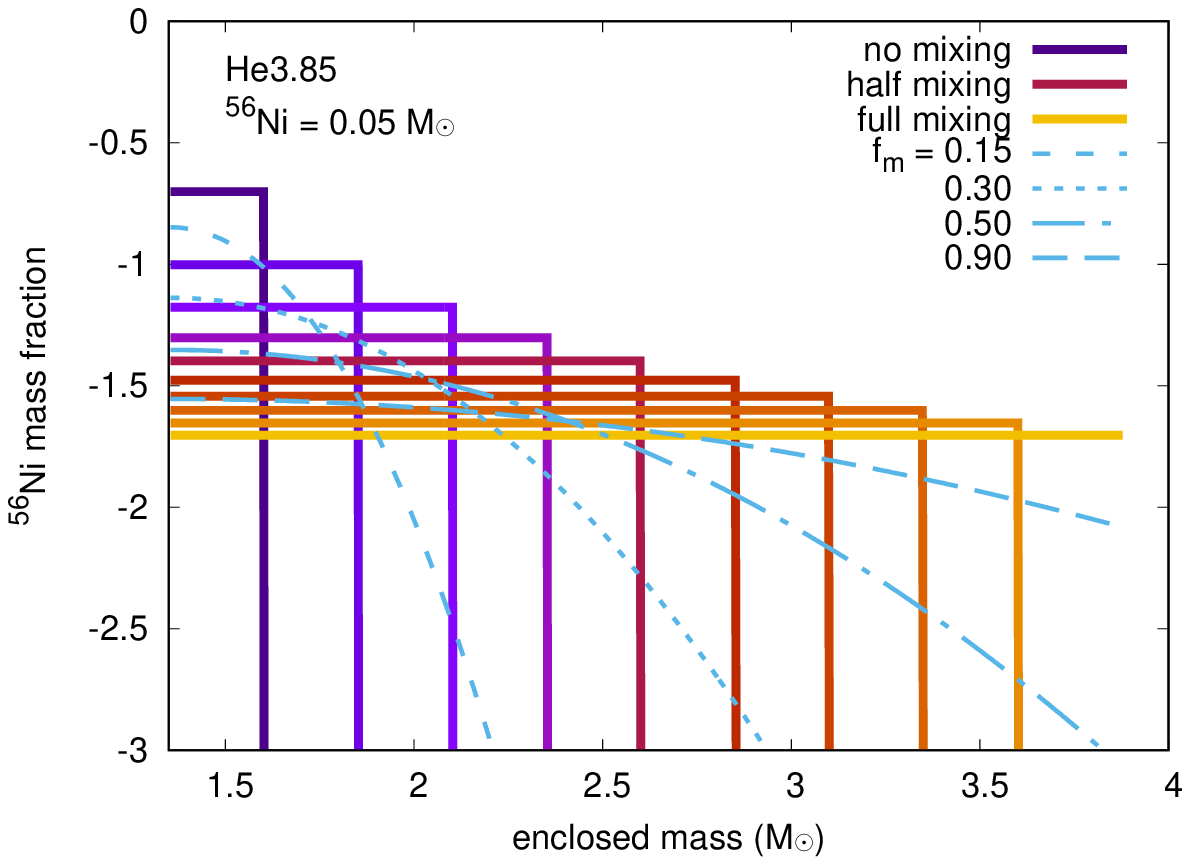}
	\includegraphics[width=\columnwidth]{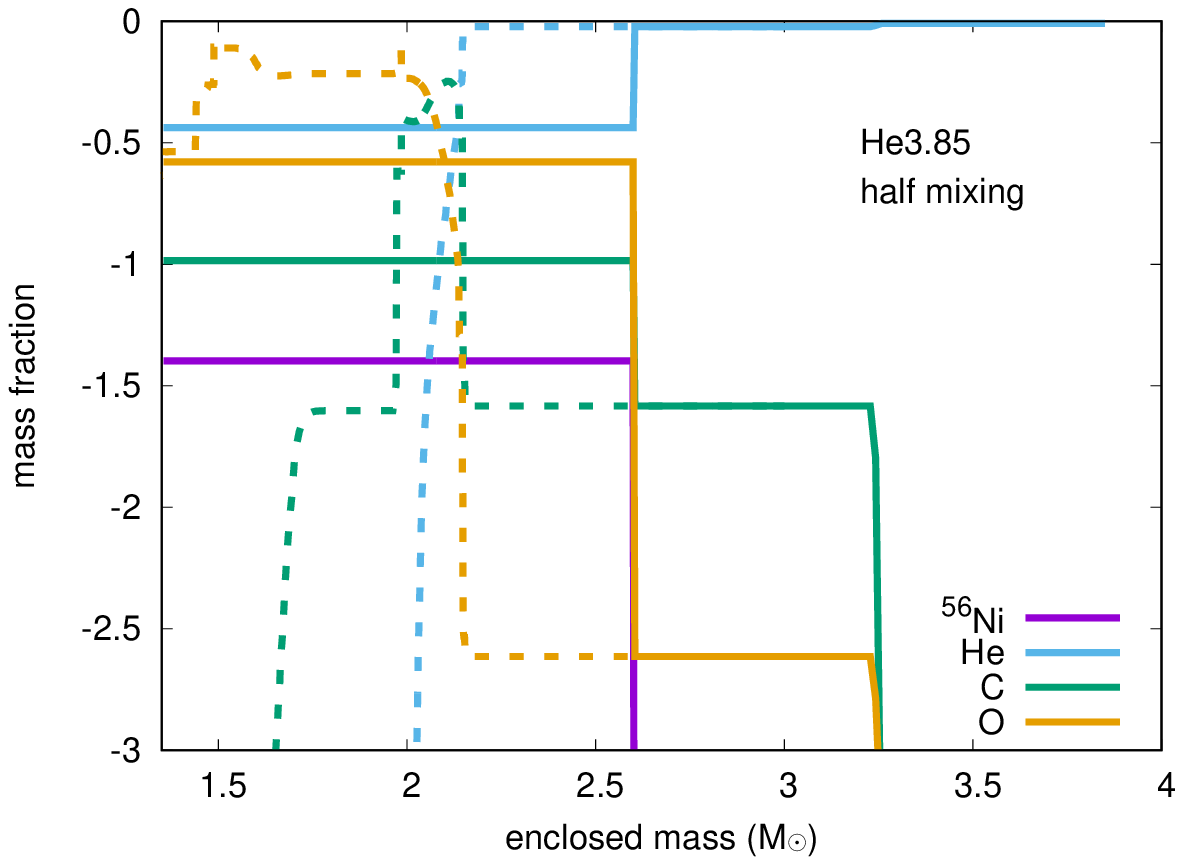}
    \caption{
    \textit{Top:} Examples of \Ni\ distributions.
    $f_m$ is the factor determining the degree of mixing defined in Eq.~(\ref{eq:yoonni}).
    \textit{Bottom:} Abundances of representative elements in the half-mixing model of He3.85 (solid lines). The original abundance profiles are shown with the dashed lines.
    }
    \label{fig:composition}
\end{figure}

\section{Introduction}
Stripped-envelope supernovae (SNe) are core-collapse SN explosions of massive stars with little or no hydrogen. Among them, Type~Ib SNe (SNe~Ib) do not have hydrogen signatures but have helium signatures in their spectra, while Type~Ic SNe (SNe~Ic) do not have signatures of both hydrogen and helium in their spectra \citep{filippenko1997}. In order not to have hydrogen signatures in spectra, the SN progenitors should contain hydrogen mass of less than 0.03~\Msun\ \citep{hachinger2012stripped}. Therefore, the progenitors of SNe~Ib and Ic are Wolf-Rayet stars that somehow lost their hydrogen-rich envelope during the evolution. Multiplicity of massive stars is suggested to play an essential role in forming stripped-envelope SN progenitors \citep[e.g.,][]{podsiadlowski1992,nomoto1995ibcprogenitor,yoon2010ibcprogenitor,eldridge2011,tauris2013} in addition to stellar wind \citep[e.g.,][]{woosley1993,meynet2005}.

Large-scale material mixing in core-collapse SN ejecta is known to occur \citep[e.g.,][]{kumagai1989}. Especially, mixing of radioactive \Ni\ strongly affects the electromagnetic properties of SNe. The effect of \Ni\ mixing in stripped-envelope SNe has long been studied \citep[e.g.,][]{ensman1988,shigeyama1990,dessart2012ibic,bersten2013,piro2013,taddia2018cspstripped,yoon2019mixing,teffs2020}. In particular, it has been suggested that the difference between SNe~Ib and SNe~Ic is in the degree of \Ni\ mixing, because non-thermal excitation required to observe the He~\textsc{i} lines can be triggered by the radioactive decay of \Ni\ \citep{lucy1991,swartz1991}. If \Ni\ is not mixed in the helium layer of the progenitor, the helium lines may not be observed even if helium exists in the progenitor. The degree of \Ni\ mixing also changes the light curves (LCs) of stripped-envelope SNe \citep[e.g.,][]{shigeyama1990}. Recently, \citet{yoon2019mixing} conducted a systematic study of the effect of \Ni\ mixing on the color evolution in stripped-envelope SNe. They found that the early color evolution can be strongly affected by the degree of \Ni\ mixing and it can be a strong probe of \Ni\ mixing. 

In this work, we investigate the effect of \Ni\ mixing on the photospheric velocity evolution of stripped-envelope SNe. Recent development in the high-cadence transient surveys started to allow us to access spectroscopic information of SNe shortly after the explosions \citep[e.g.,][]{gal-yam2014flash,yaron2017flash}. Velocity information that can be acquired through spectra has long been used to obtain an independent constraint on stripped-envelope SN properties \citep[e.g.,][]{iwamoto2000,mazzali2000,lyman2016,dessart2016,taddia2018cspstripped,prentice2019}. Indeed, the previous studies on \Ni\ mixing in stripped-envelope SNe indicate that \Ni\ mixing can affect the photospheric velocity evolution in stripped-envelope SNe \citep[e.g.,][]{dessart2012ibic}. In this paper, we perform a systematic study of the effect of \Ni\ mixing on the photospheric velocity evolution in stripped-envelope SNe.

The rest of this paper is organized as follows. We summarize how we get progenitor models, how we mix \Ni, and how we conduct radiation transfer simulations in Section~\ref{sec:methods}. Our results are presented in Section~\ref{sec:results}. We discuss the results and conclude this paper in Section~\ref{sec:discussion}. 

\begin{figure}
	\includegraphics[width=\columnwidth]{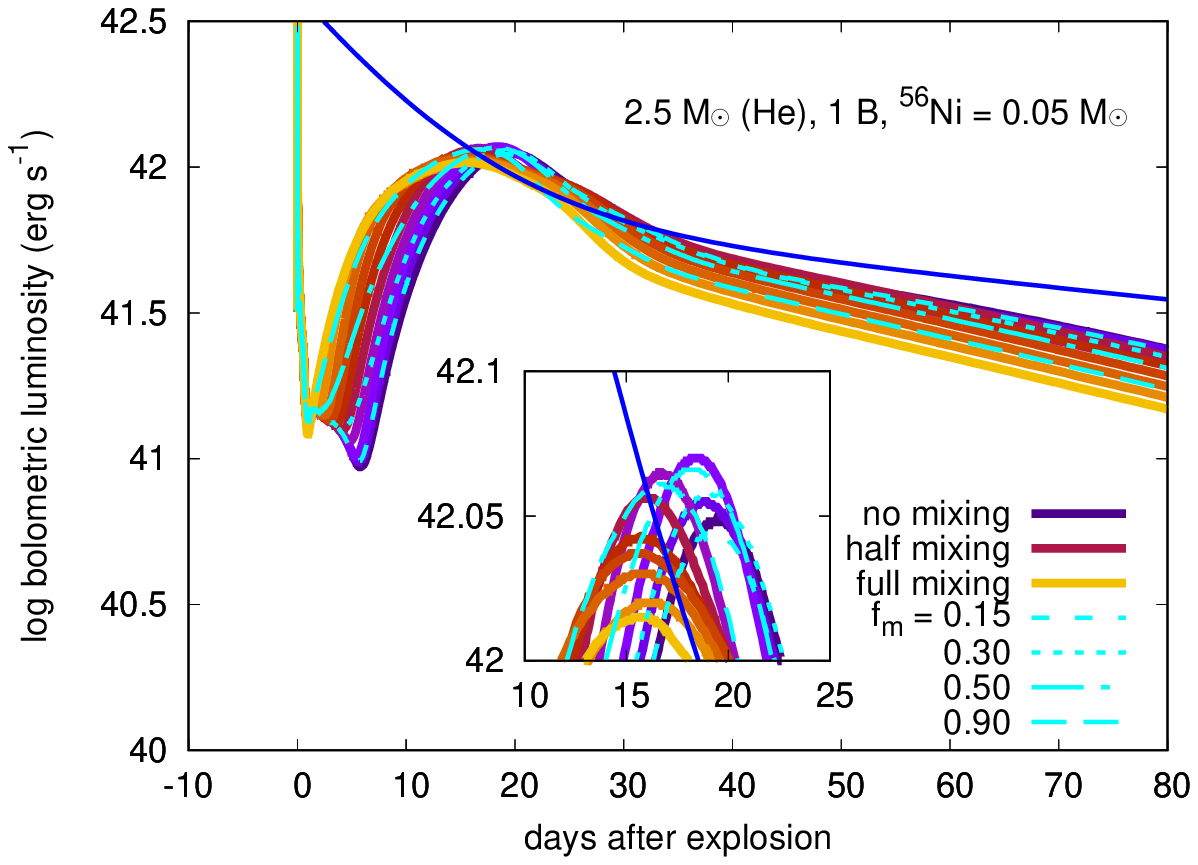}
	\includegraphics[width=\columnwidth]{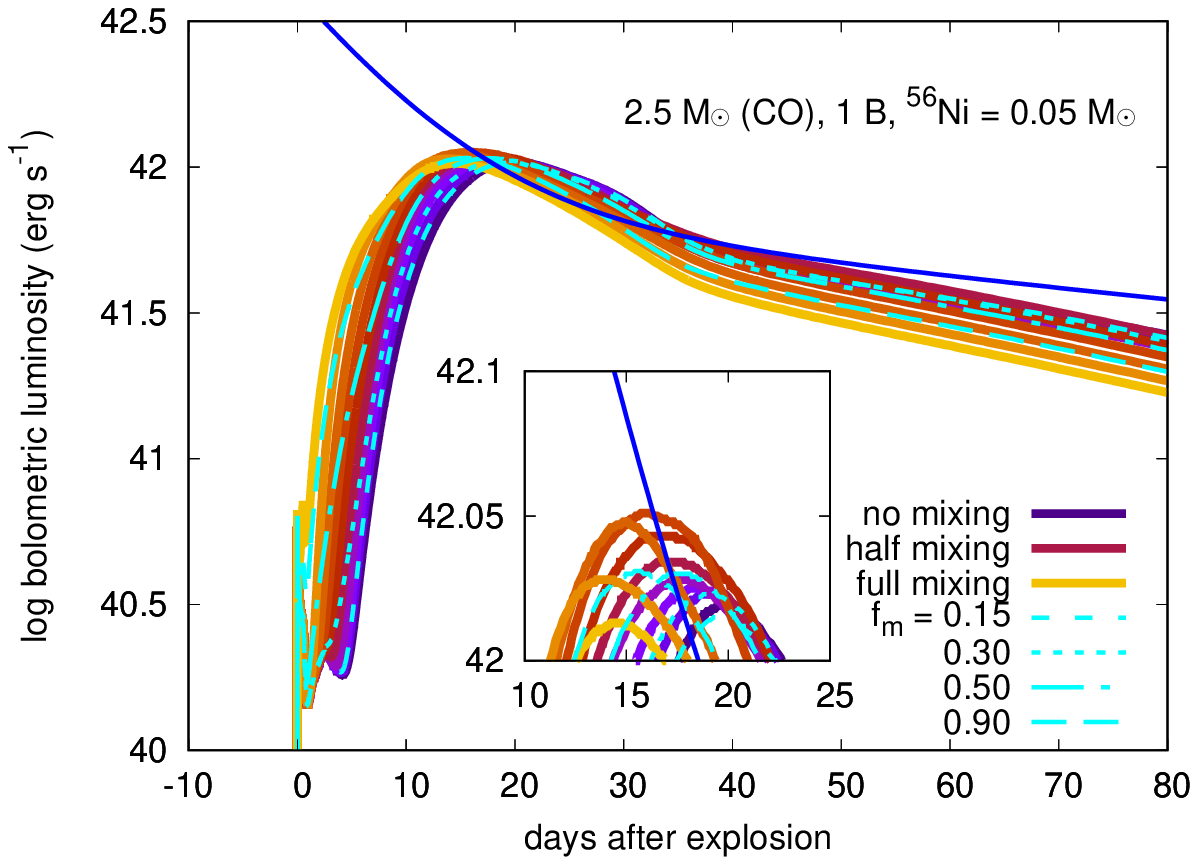}
    \caption{
    Bolometric LCs with different \Ni\ mixing from the helium progenitor (top) and the carbon+oxygen progenitor (bottom). The models have the ejecta mass of 2.5~\Msun, the explosion energy of 1~B, and the \Ni\ mass of 0.05~\Msun. The blue lines are the energy deposition rate of the $\Ni\rightarrow {}\Co\rightarrow {}\Fe$ decay from 0.05~\Msun\ of \Ni. The insets zoom in the peaks of the LCs.
    }
    \label{fig:m2p5_1b_lc}
\end{figure}

\begin{figure}
	\includegraphics[width=\columnwidth]{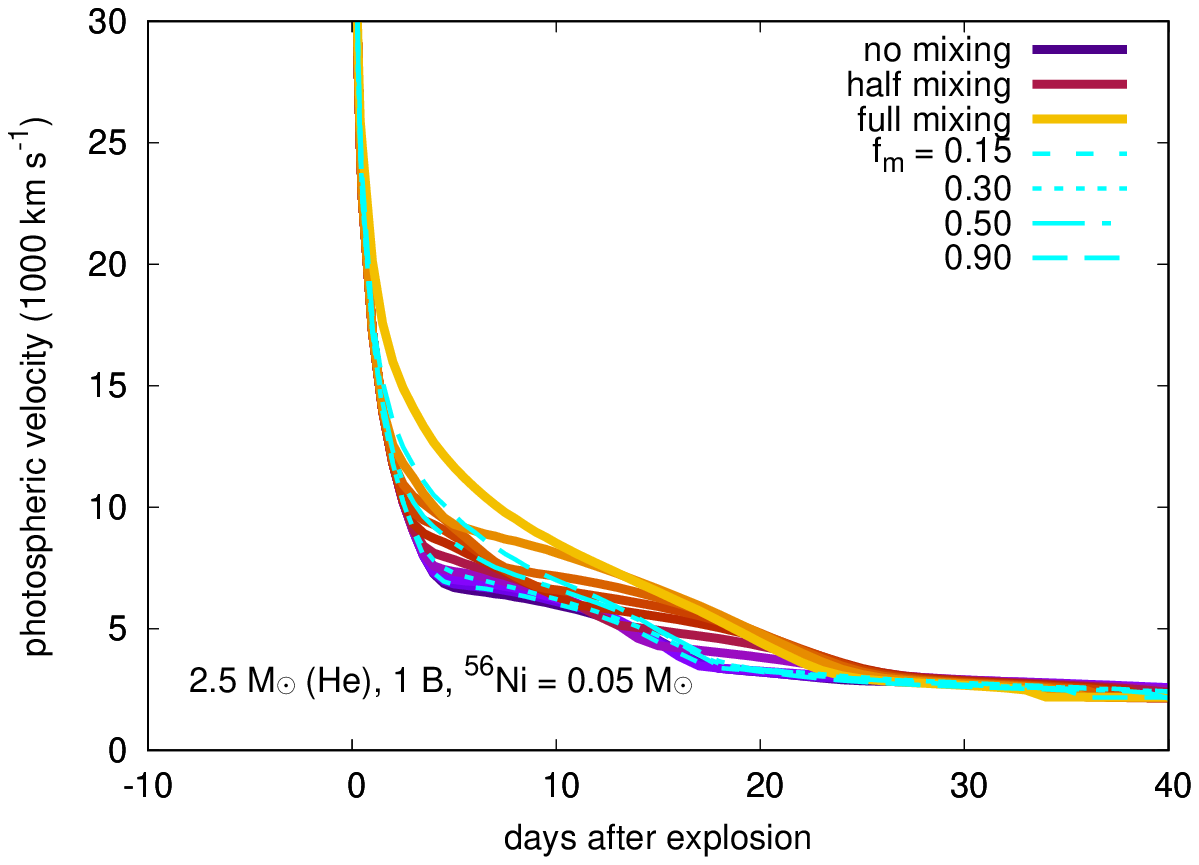}
	\includegraphics[width=\columnwidth]{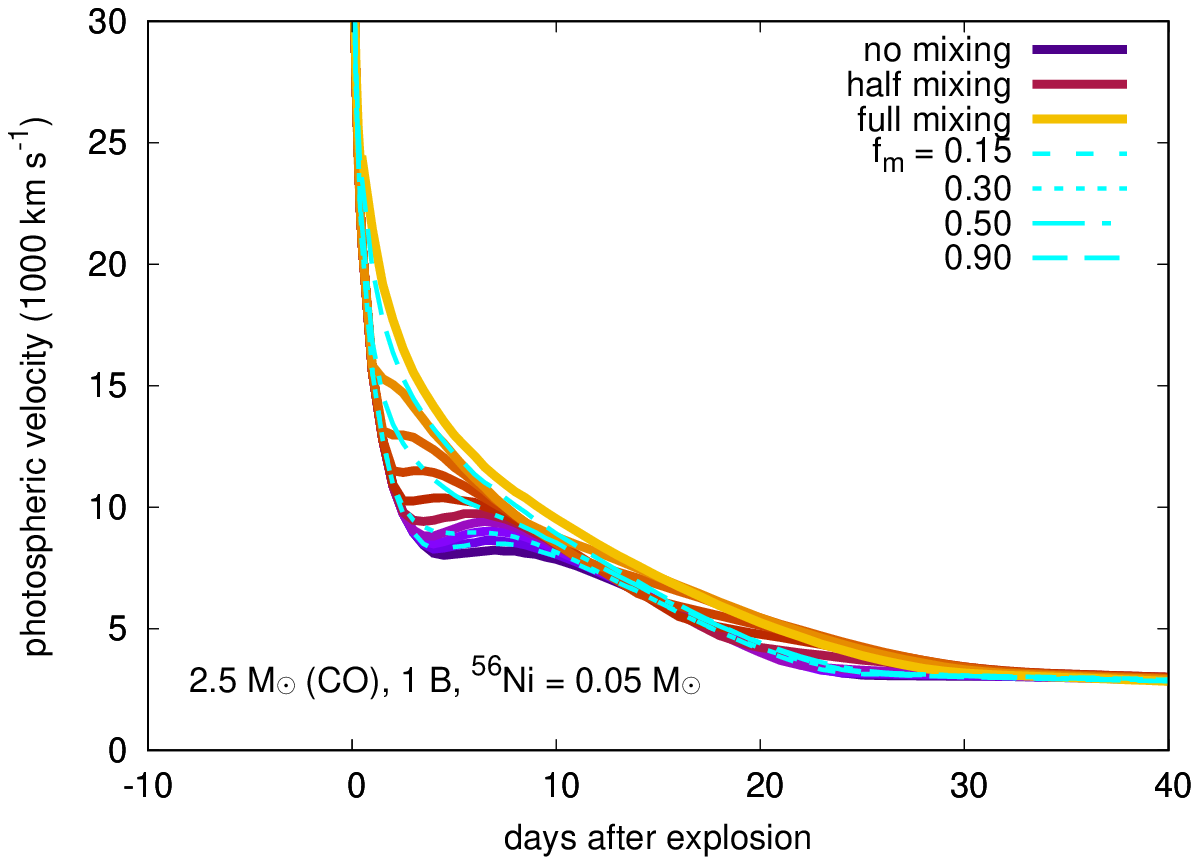}
    \caption{
    Photospheric velocity evolution of the models presented in Fig.~\ref{fig:m2p5_1b_lc}.
    }
    \label{fig:m2p5_1b_vel}
\end{figure}

\begin{figure}
	\includegraphics[width=\columnwidth]{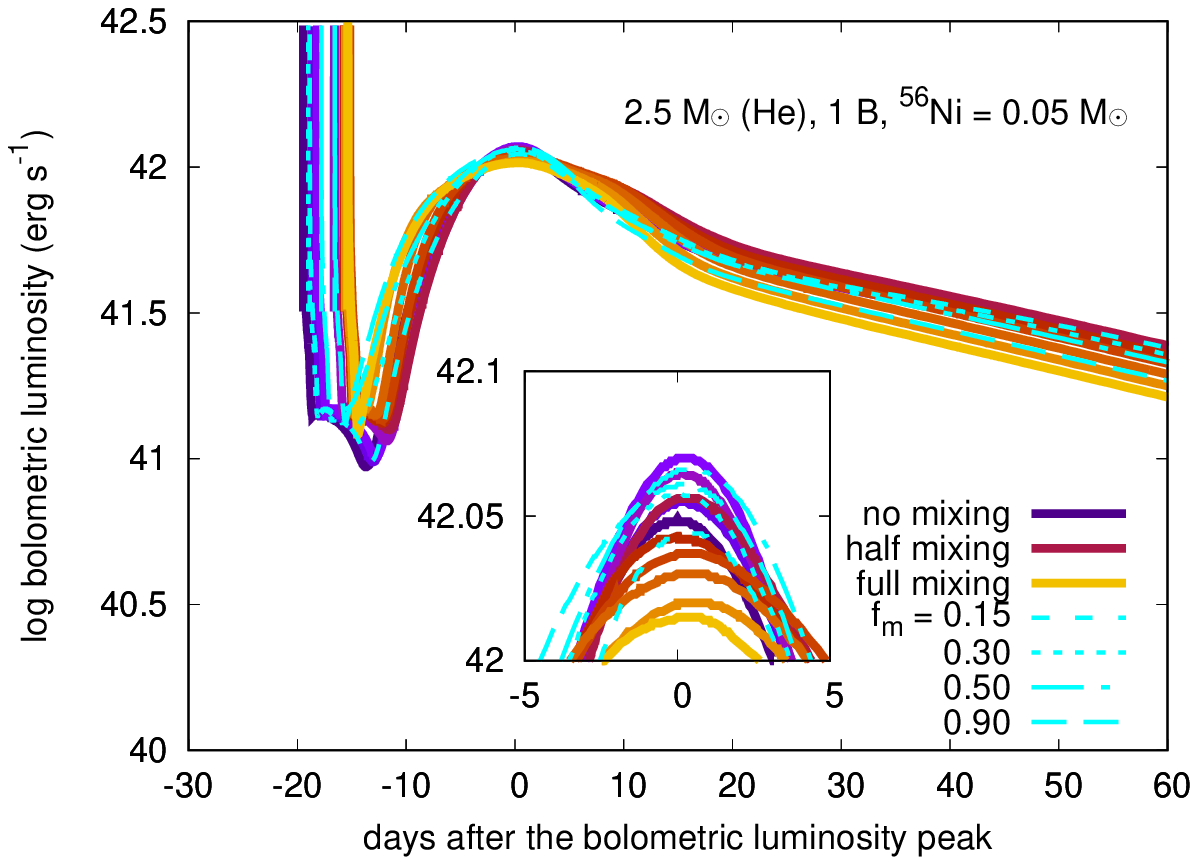}
	\includegraphics[width=\columnwidth]{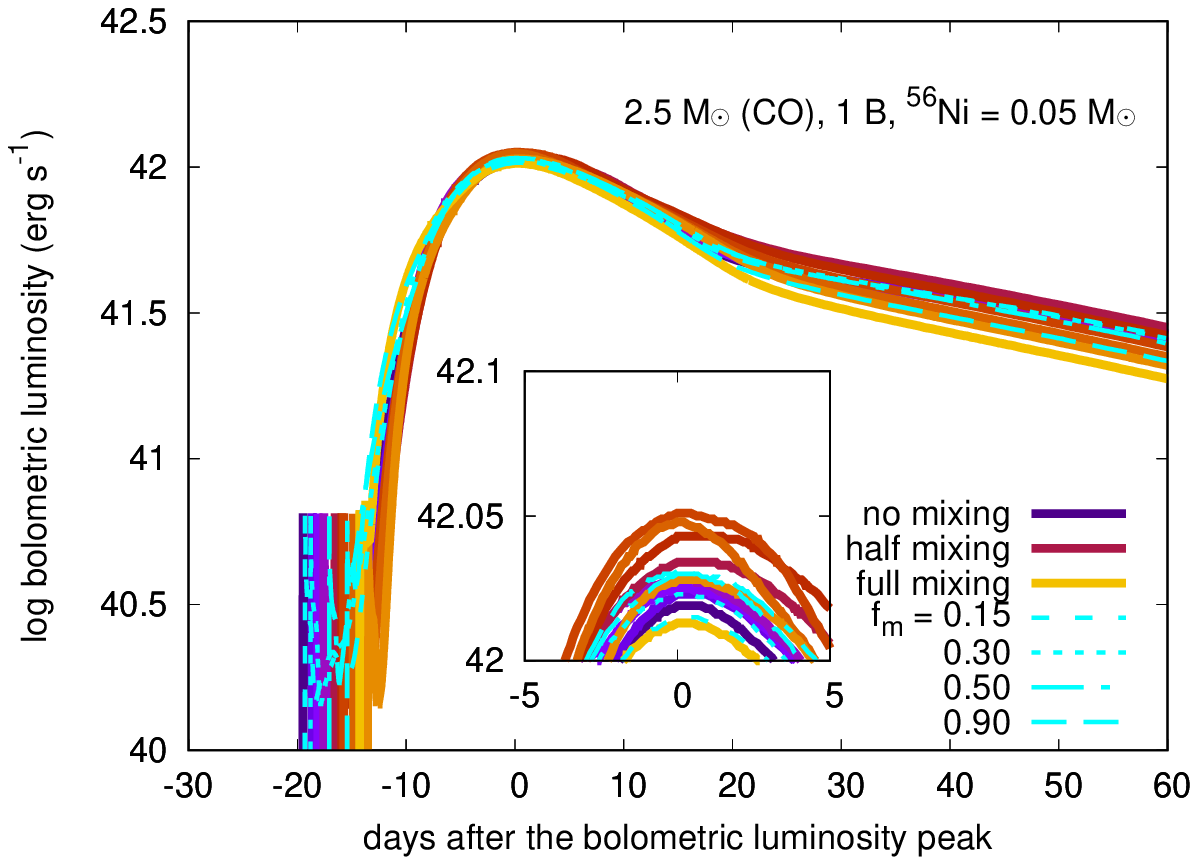}
    \caption{
    Same as Fig.~\ref{fig:m2p5_1b_lc}, but the origin of the time axis is set at the luminosity peak after the shock breakout.
    The insets zoom in the peaks of the LCs.
    }
    \label{fig:m2p5_1b_lc_peak}
\end{figure}

\begin{figure}
	\includegraphics[width=\columnwidth]{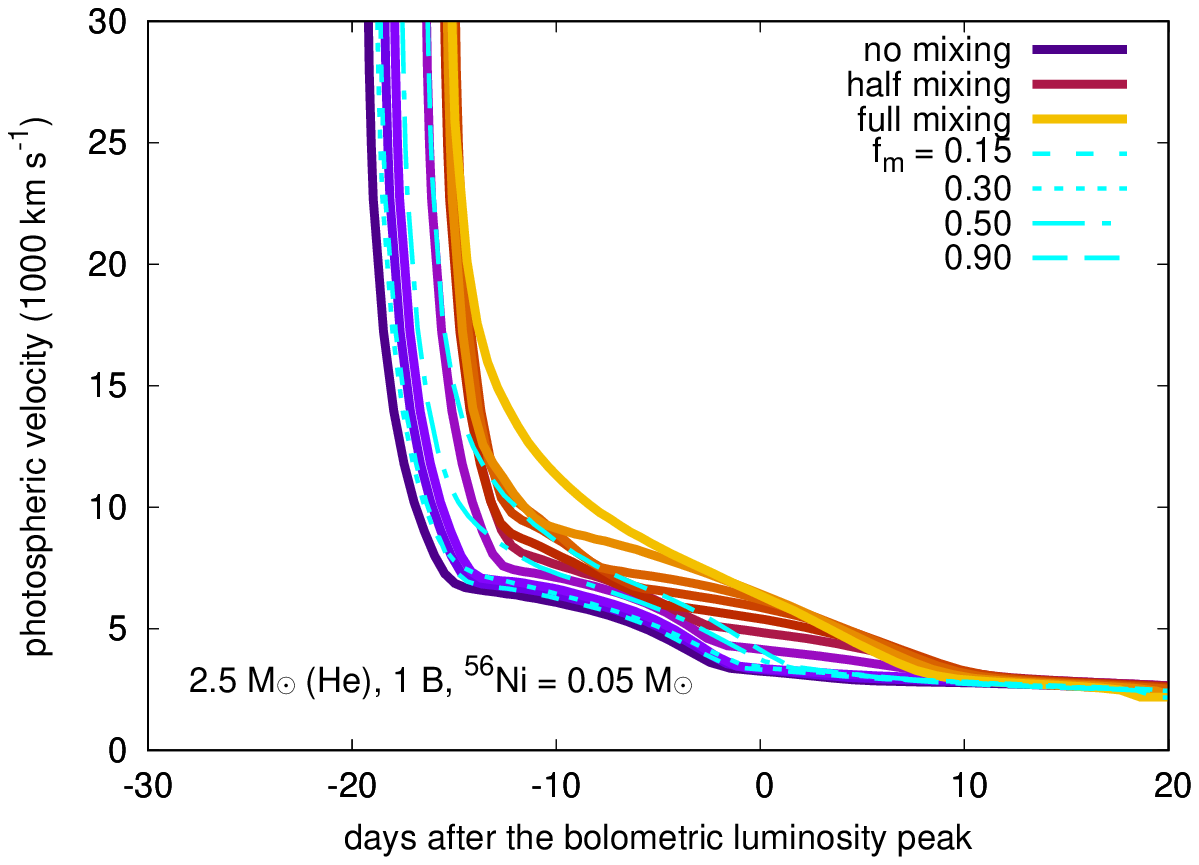}
	\includegraphics[width=\columnwidth]{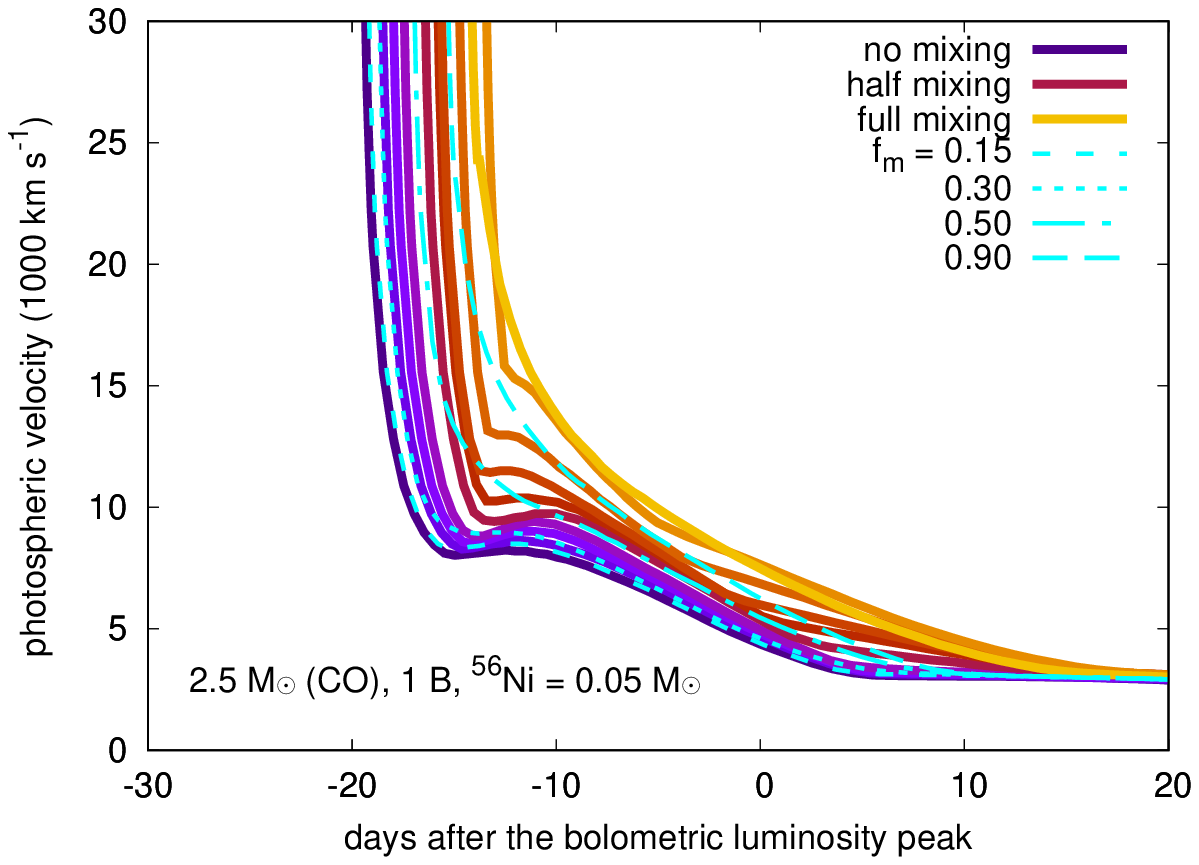}
    \caption{
    Same as Fig.~\ref{fig:m2p5_1b_vel}, but the origin of the time axis is set at the luminosity peak after the shock breakout.
    }
    \label{fig:m2p5_1b_vel_peak}
\end{figure}

\section{Methods}\label{sec:methods}
\subsection{Progenitor models}
We obtain hydrogen-free progenitor models by using the public stellar evolution code \texttt{MESA} \citep{paxton2011,paxton2013,paxton2015,paxton2018,paxton2019}. We assume the metallicity  of $Z=0.02$. We adopt the standard mixing-length theory for convection with the mixing-length parameter of 2.0 with the Schwarzschild criteria. 

Our focus in this paper is on hydrogen-free SNe. There are many mechanisms to remove the hydrogen-rich envelope from SN progenitors and the mass loss mechanisms themselves are a largely investigated field of research \citep{smith2014ARA&A}. In this work, we artificially remove hydrogen-rich envelopes to obtain hydrogen-free SN progenitors. We first evolve massive stars from zero-age main sequence (ZAMS) to the end of helium burning without mass loss. The helium core mass does not change much after the helium burning. At this point, we stop the stellar evolution calculation and gradually reduce the mass from the surface through wind by using the \texttt{relax\_mass} option in \texttt{MESA}. The maximum mass-loss rate allowed in this process is $10^{-4}~\Msunpyr$ (\texttt{lg\_max\_abs\_mdot = -4} in \texttt{MESA}). When the stellar mass reaches a desired mass, we resume the stellar evolution calculation without mass loss. The stellar evolution calculations are performed at least until the end of oxygen burning. The subsequent evolution would not affect the envelope structure of the progenitors significantly.

Using the method described above, we obtain three hydrogen-free progenitor models; He3.85, CO3.85, and He4.90 (Table~\ref{tab:progenitors}). The density structure of the progenitors is shown in Fig.~\ref{fig:density}. He3.85 and CO3.85 have the same mass (3.85~\Msun), but He3.85 is a helium star while CO3.85 is a carbon+oxygen star without helium. The ZAMS masses of He3.85 and CO3.85 are 14~\Msun\ and 20~\Msun, respectively. He4.90 is a 4.90~\Msun\ helium star progenitor with the ZAMS mass of 17~\Msun. These masses are determined to have the typical ejecta mass ($\simeq 2~\Msun$) in stripped-envelope SNe \citep[e.g.,][]{lyman2016}.

When we explode these progenitors, we set a mass cut $M_\mathrm{cut}$ below which is assumed to form a compact remnant and is not ejected. The mass cut is set at 1.35~\Msun\ (He3.85 and CO3.85) and 1.40~\Msun\ (He4.90) so that the ejecta mass becomes 2.5~\Msun\ (He3.85 and CO3.85) and 3.5~\Msun\ (He4.90), respectively.

\subsection{\Ni\ mixing}
Using the progenitor models presented so far, we artificially put \Ni\ in the ejecta with many different degrees of mixing. The ``full mixing'' model is the model in which \Ni\ is uniformly mixed in the entire ejecta. From the full mixing model, we reduce the outer edge of the layer where \Ni\ is mixed in steps of 0.25~\Msun\ as shown in Fig.~\ref{fig:composition}. The model with the least degree of mixing, in which \Ni\ exists up to 0.25~\Msun\ above the mass cut, is labeled as ``no mixing.'' We mix all the elements in the region where we include \Ni\ (Fig.~\ref{fig:composition}). We decrease the mass of the other elements following the original mass fraction and replace it with \Ni. 

To compare our results with those of the previous study by \citet{yoon2019mixing}, we also adopt their mixing method. \citet{yoon2019mixing} use the \Ni\ distribution in the ejecta formulated as
\begin{equation}
    X_{\Ni}(M_r) = A\exp\left(-\left[\frac{M_r-M_\mathrm{cut}}{f_m\left(M_\mathrm{tot}-M_\mathrm{cut}\right)}\right]^2\right), \label{eq:yoonni}
\end{equation}
where $M_r$ is the mass coordinate, $A$ is the scaling factor, $M_\mathrm{tot}$ is the total mass, and $f_m$ is the factor determining the degree of mixing. As in \citet{yoon2019mixing}, we adopt $f_m = 0.15, 0.30, 0.50$ and 0.90. Fig.~\ref{fig:composition} shows examples of the \Ni\ distribution with these $f_m$. When adopting this \Ni\ distribution, we just reduce the mass of the other elements following the mass fraction to put \Ni\ and no other elements are mixed.

Our standard models have the \Ni\ mass of 0.05~\Msun. This mass is chosen based on the recent study of well-observed stripped-envelope SNe (\citealt{meza2020ni}, see also \citealt{anderson2019ni}). We also investigate models with a different \Ni\ mass (0.1~\Msun).

\subsection{Light-curve calculations}
LC calculations are performed by using the one-dimensional multi-group radiation hydrodynamics code \texttt{STELLA} \citep{blinnikov1998sn1993j,blinnikov2000sn1987a,blinnikov2006sniadeflg}. 
\texttt{STELLA} calculates the spectral energy distributions (SEDs) at each time-step and obtains multicolor LCs by convolving filter functions with the SEDs. Bolometric LCs are obtained by integrating the SEDs. \texttt{STELLA} implicitly treats time-dependent equations of the angular moments of intensity averaged over a frequency bin with the variable Eddington method in addition to hydrodynamics. The SN explosions are artificially triggered by putting thermal energy just above the mass cut. The photosphere is defined as where the Rosseland mean optical depth becomes $2/3$.

We use the non-relativistic version of \texttt{STELLA}. However, the outermost layers in the ejecta after the shock breakout becomes close to the speed of light because our progenitors are compact. To avoid the numerical difficulties caused by the regions close to the speed of light, we remove the region having velocity exceeding 100,000~\kmps\ shortly after the shock breakout. The mass of the removed region depends on the models, but they are less than 0.003~\Msun. The synthetic LCs at shock breakout and immediately after are affected by the mass removal, but the later LCs would not be affected because of the small removed mass. The shock breakout signals in our models are, therefore, not calculated properly but they are irrelevant to this study and should be ignored.

\begin{figure}
	\includegraphics[width=\columnwidth]{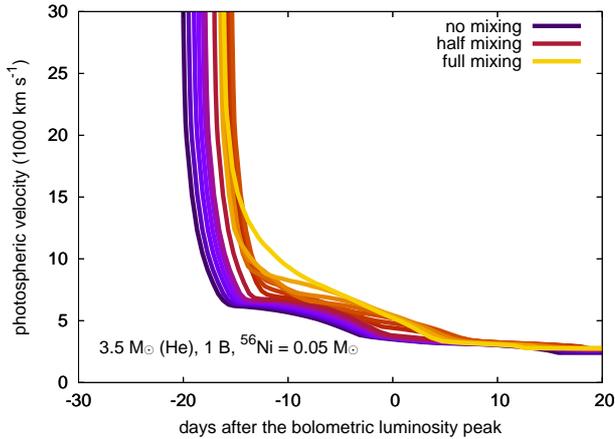}
    \caption{
    Photospheric velocity evolution for the models having the ejecta mass of 3.5~\Msun.
    }
    \label{fig:m3p5he_1b_ni0p05}
\end{figure}

\section{Results}\label{sec:results}
\subsection{Bolometric LCs}
Fig.~\ref{fig:m2p5_1b_lc} shows the synthetic bolometric LCs with different degrees of \Ni\ mixing from our helium and carbon+oxygen progenitors with $\Mej=2.5~\Msun$. The models in the figure assume the explosion energy of $1~\mathrm{B} \equiv 10^{51}~\mathrm{erg}$.

We first look into the helium star explosion models. After the shock breakout, the bolometric luminosity declines quickly until the recombination front appears in the ejecta. After this moment, the photosphere is kept at the recombination front for a while and the luminosity decline rate becomes significantly smaller than before \citep[e.g.,][]{ensman1988,dessart2011,bersten2013}.
Not only helium but also other elements such as carbon and oxygen are related in the recombination. The photospheric temperature evolution during the recombination phase is similar to those found in the previous studies (Section~\ref{sec:color}).
The bolometric LCs start to increase when the heating caused by the \Ni\ decay becomes large enough to push the recombination front outwards. The temperature of the ejecta increases again and the subsequent LC evolution follows the standard LC evolution of stripped-envelope SNe. The peak luminosity roughly matches the nuclear energy deposition rate (the blue lines in Fig.~\ref{fig:m2p5_1b_lc}) at the time of the peak luminosity \citep{arnett1982,blinnikov2006sniadeflg,khatami2019kasen}.
The peak bolometric luminosity does not strongly depend on the degree of \Ni\ mixing, but there are slight differences likely caused by the difference in the gamma-ray trapping efficiency due to the different degrees of mixing. It is natural, because, for a given amount of radioactive material, higher degree of mixing implies more of the material closer to the surface, hence, less gamma-photons to be trapped and reprocessed to produce the LC.

The major differences in the bolometric LCs caused by \Ni\ mixing are found in two phases. The first phase is the transition from the recombination phase to the luminosity increase triggered by heating from the nuclear decay. The recombination phase becomes shorter with the larger degree of mixing. When \Ni\ only locates at the center, the heating of ejecta only occurs at the center and temperature in the ejecta gradually increases from inside. On the other hand, when \Ni\ is mixed, the heating by the nuclear decay in the outer layers in the ejecta becomes efficient and the increase in the ejecta temperature occurs earlier. This makes the recombination phase shorter for the more mixed models. The efficient heating also makes the rise time of the bolometric LC shorter and thus more mixed models have shorter rise times. The second phase that is strongly affected by the degree of \Ni\ mixing is the tail phase. More mixed models have a less tail luminosity because gamma-rays from the nuclear decay can be less trapped in them \citep[e.g.,][]{wheeler2015,sharon2020}.

The behavior of the carbon+oxygen star explosion models is basically the same as that of the helium star explosion models. The recombination phase after the shock breakout is determined by the recombination temperature without helium. The luminosity at the recombination phase is much lower than that of the helium star explosions. The bolometric LC evolution after the recombination phase is similar to that of the helium star explosions.

\begin{figure}
    \includegraphics[width=\columnwidth]{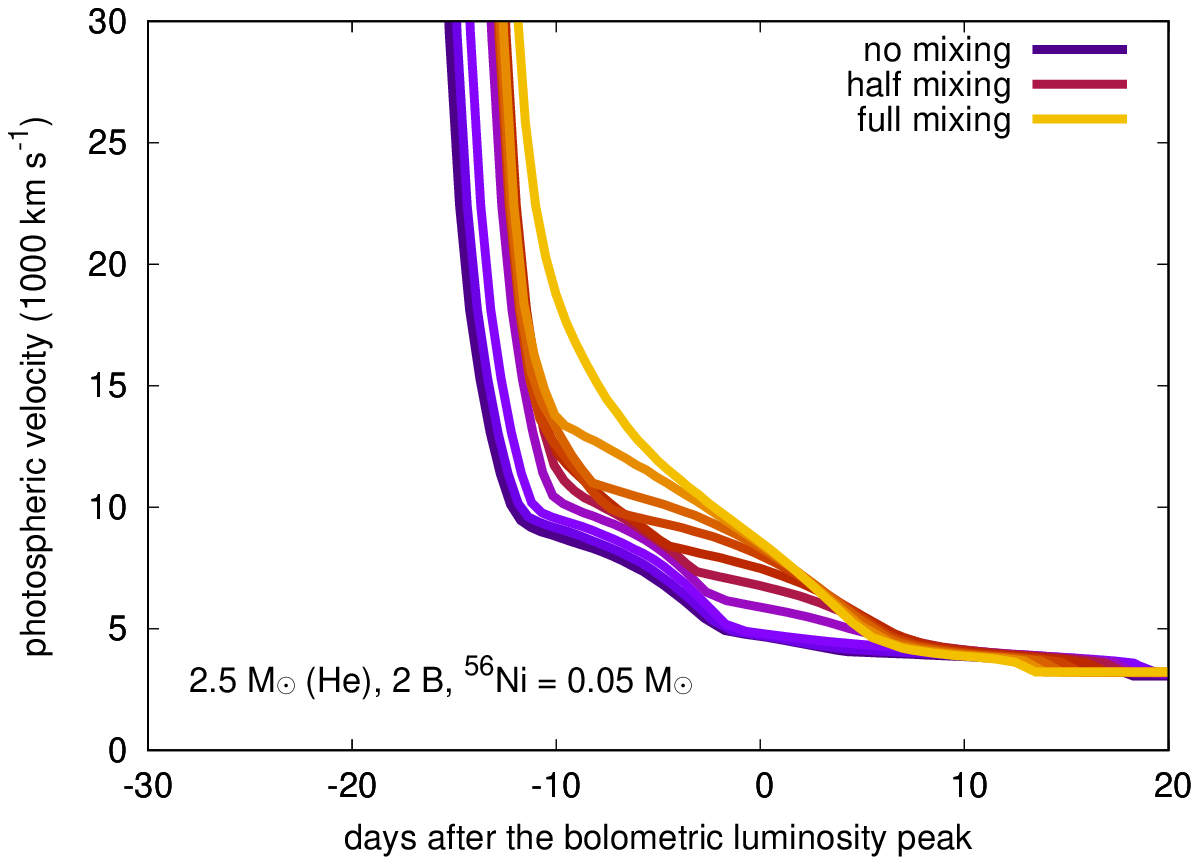}   
    \includegraphics[width=\columnwidth]{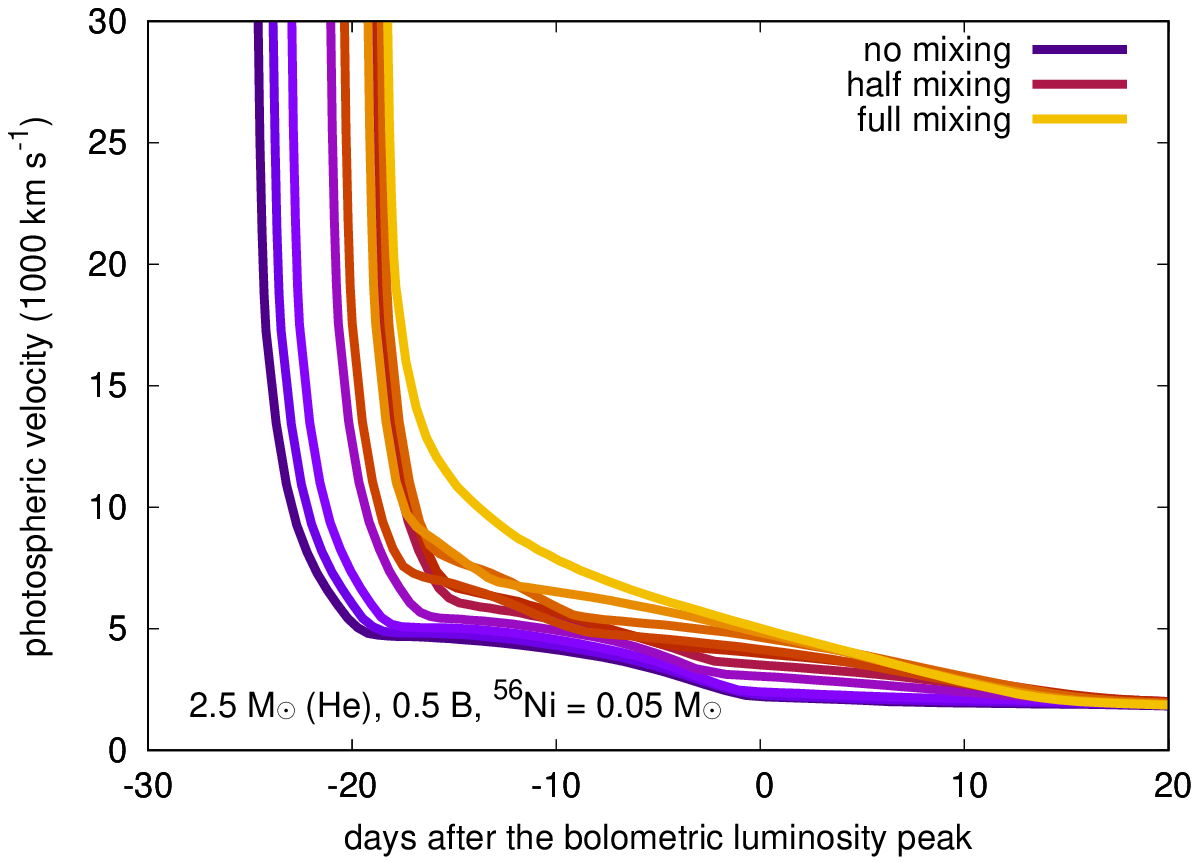}       
    \caption{
    Photospheric velocity evolution for the models having the explosion energy of 2~B (top) and 0.5~B (bottom).
    }
    \label{fig:m2p5he_?b_vel_peak}
\end{figure}

\subsection{Photospheric velocity}
We now look into the photospheric velocity.
Fig.~\ref{fig:m2p5_1b_vel} shows the photospheric velocity evolution of the LC models discussed in the previous section. We can see that the photospheric velocity evolution is strongly affected by the degree of \Ni\ mixing. After the shock breakout, the photospheric velocity quickly recedes inwards as the ejecta cool adiabatically. The photospheric velocity evolution suddenly flattens when the recombination phase ends and the bolometric luminosity starts to increase due to the heating from the nuclear decay. The time when the photospheric velocity evolution flattens depends on the degree of mixing and it can be used to observationally constrain \Ni\ mixing in stripped-envelope SNe.

We have so far investigated the models with the step-function \Ni\ distribution. The border of the \Ni\ mixing is not necessarily defined sharply. \citet{yoon2019mixing} adopted exponential \Ni\ distribution (Eq.~\ref{eq:yoonni}) and we show the models with their \Ni\ distribution in Figs.~\ref{fig:m2p5_1b_lc} and \ref{fig:m2p5_1b_vel}. We find that the effect of the \Ni\ mixing found by the step-function distribution remains even when we adopt the exponential \Ni\ distribution from \citet{yoon2019mixing}.

Figs.~\ref{fig:m2p5_1b_lc_peak} and \ref{fig:m2p5_1b_vel_peak} show the same models as presented in Figs.~\ref{fig:m2p5_1b_lc} and \ref{fig:m2p5_1b_vel}, but the origin of the time is set at the time of the peak luminosity after the shock breakout. Looking at the bolometric LC evolution of the helium star progenitor, we find that it takes more time after the explosion to reach the maximum bolometric luminosity in the less mixed models. However, the less mixed models have the longer recombination phase. When we look at the time it takes to reach the peak luminosity from the end of the recombination phase, the less mixed models tend to have shorter time to reach the peak luminosity. Thus, the LCs of the less mixed models rise quicker after the recombination phase. We can also see this in the photospheric velocity evolution -- the less mixed models start to have flat velocity evolution longer before the LC peak.

It needs to be stressed that the photospheric velocity at the LC peak differs significantly just by changing \Ni\ mixing without changing the ejecta mass and explosion energy. The photospheric velocity is often used to constrain $(\Eej/\Mej)^{0.5}$ in stripped-envelope SNe. However, the models presented so far that have the same $\Eej=1~\mathrm{B}$ and $\Mej=2.5~\Msun$ show significant differences in the photospheric velocity by only changing \Ni\ mixing.
We often find about a factor of 1.5 difference in the photospheric velocity at the luminosity peak in the most and least mixed models. In the analytic model, the ejecta mass and energy are scaled as $\Mej \propto v_\mathrm{ph}$ and $\Eej \propto v_\mathrm{ph}^3$, respectively, where $v_\mathrm{ph}$ is the photospheric velocity \citep[e.g.,][]{lyman2016}. Thus, the difference in the factor of 1.5 changes the \Mej\ estimate by a factor of 1.5 and \Eej\ estimate by a factor of 3.4.
The mixing effect needs to be taken into account when estimating the ejecta properties in stripped-envelope SNe.

\begin{figure}
	\includegraphics[width=\columnwidth]{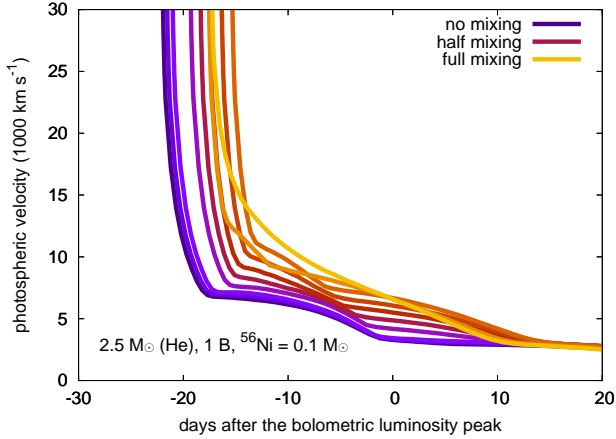}
    \caption{
    Photospheric velocity evolution for the models having the \Ni\ mass of 0.1~\Msun.
    }
    \label{fig:m2p5he_1b_ni0p1}
\end{figure}

The models presented so far has the ejecta mass of 2.5~\Msun, the explosion energy of 1~B, and the \Ni\ mass of 0.05~\Msun. We present the photospheric velocity evolution of the models with a larger ejecta mass (3.5~\Msun) in Fig.~\ref{fig:m3p5he_1b_ni0p05}, two different explosion energies (2~B and 0.5~B) in Fig.~\ref{fig:m2p5he_?b_vel_peak}, and a lager \Ni\ mass (0.1~\Msun) in Fig.~\ref{fig:m2p5he_1b_ni0p1}. We only show the models with helium star progenitors but the trends do not differ significantly even if we take the carbon+oxygen star progenitor. The effect of the \Ni\ mixing on the photospheric velocity evolution discussed so far remains unchanged.

\begin{figure}
	\includegraphics[width=\columnwidth]{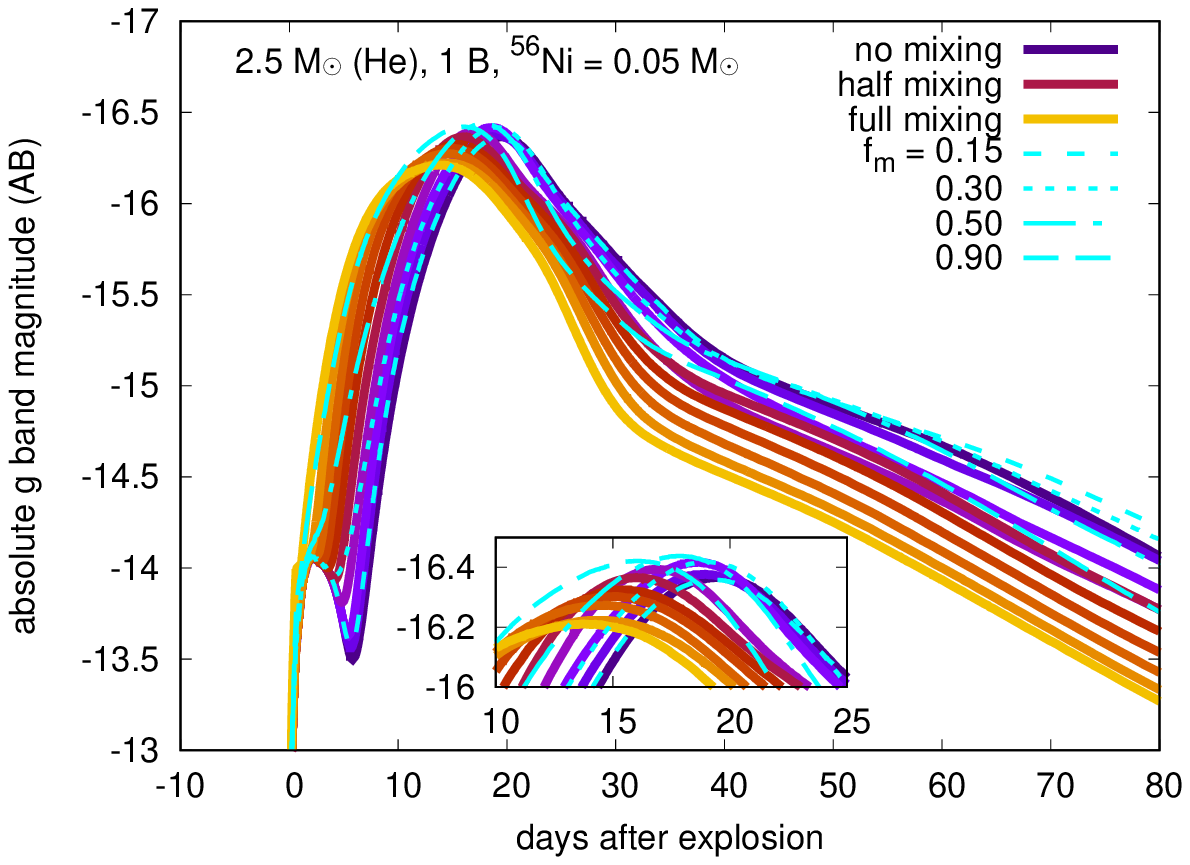}
	\includegraphics[width=\columnwidth]{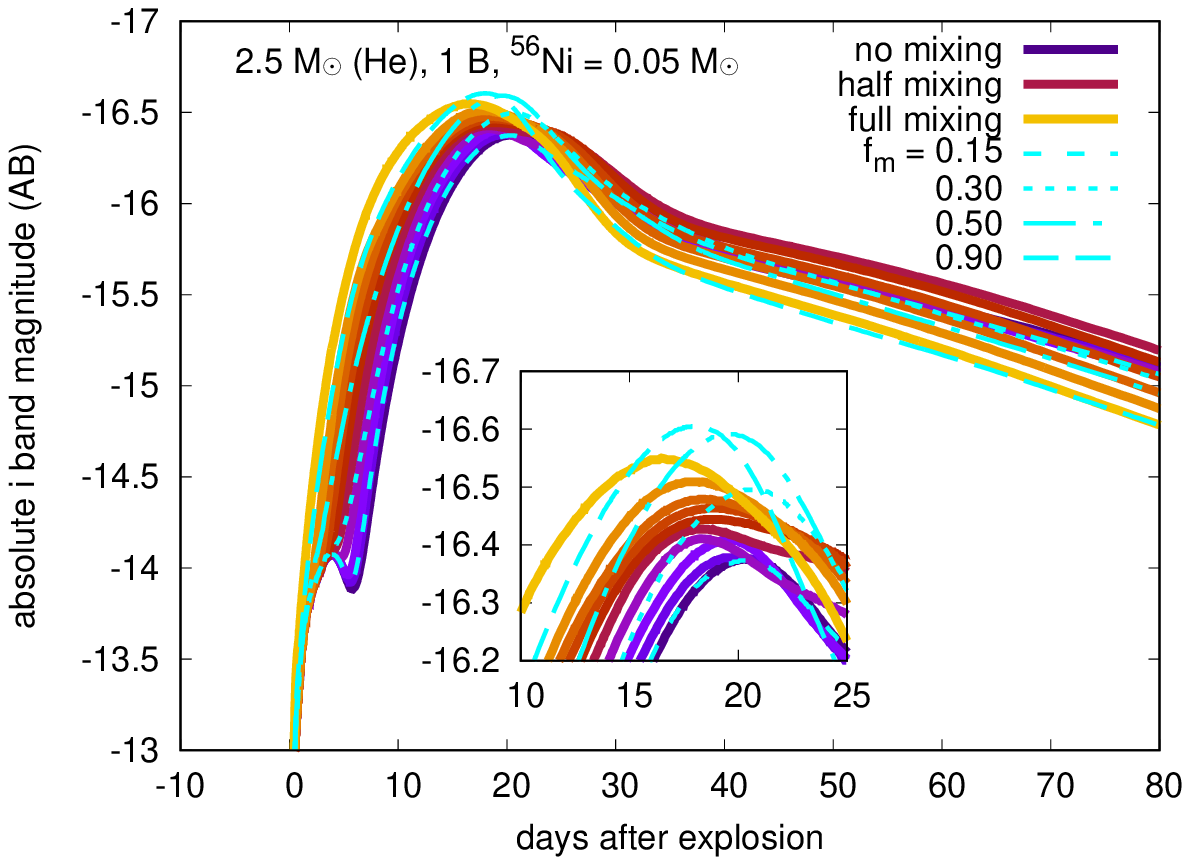}
    \caption{
    $g$ (top) and $i$ (bottom) band LCs for the helium star explosion models.
    The insets zoom in the peaks of the LCs.
    }
   \label{fig:m2p5he_1b_gi}
\end{figure}

\begin{figure}
	\includegraphics[width=\columnwidth]{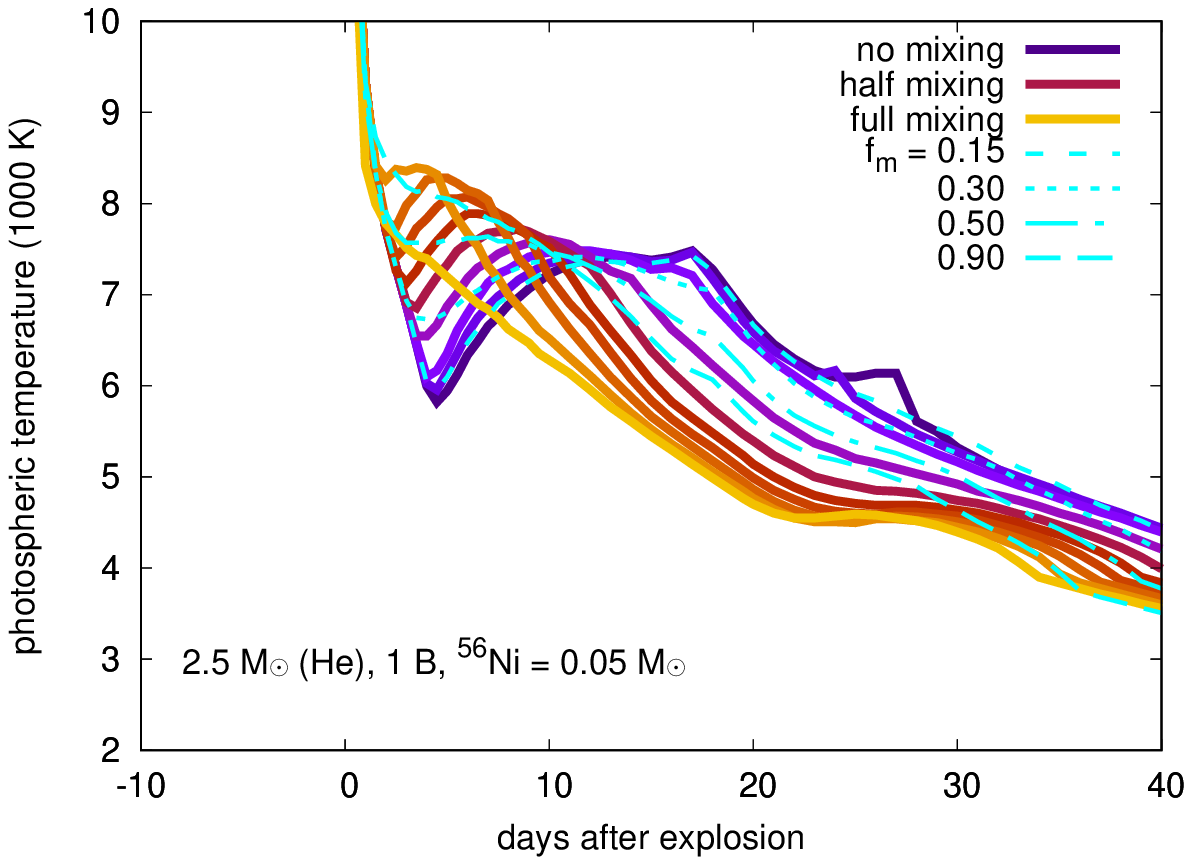}
	\includegraphics[width=\columnwidth]{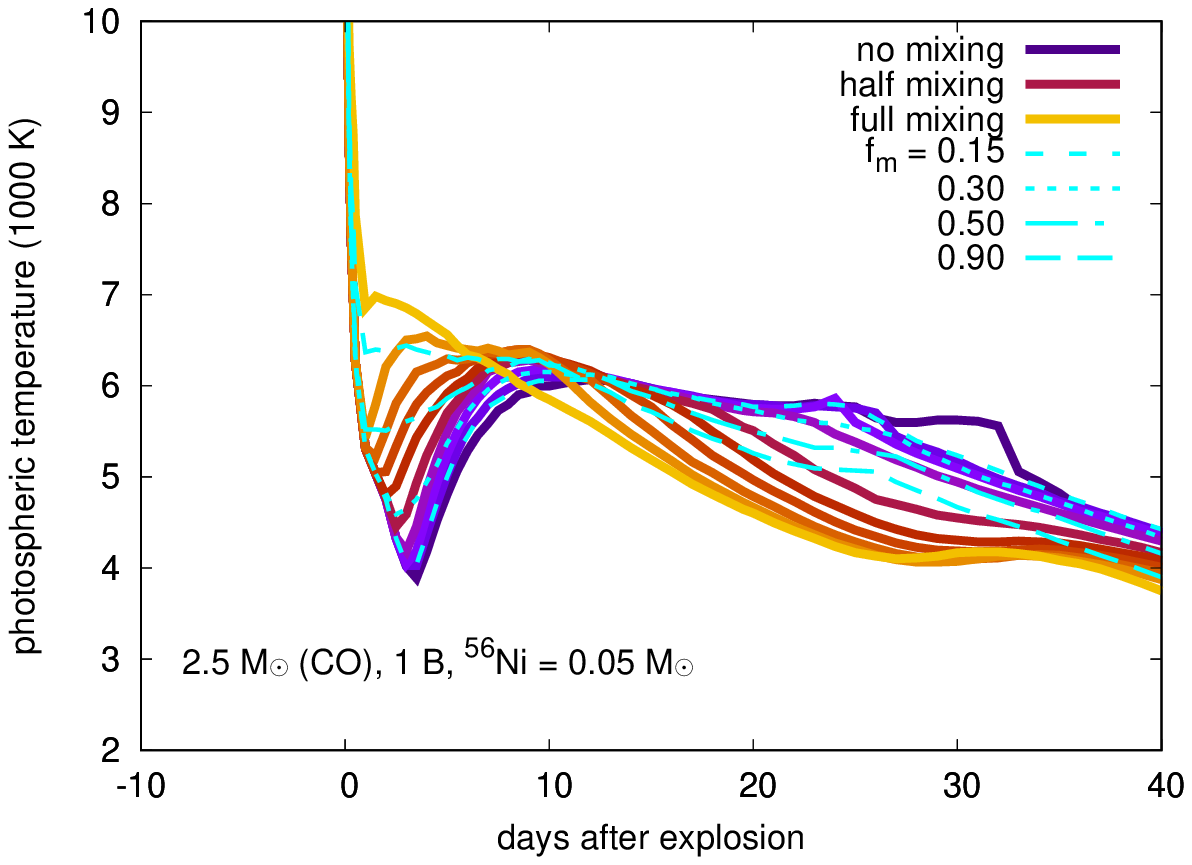}
    \caption{
    Photospheric temperature evolution of the models presented in Fig.~\ref{fig:m2p5_1b_lc}.
    }
    \label{fig:m2p5_1b_temp}
\end{figure}

\begin{figure}
	\includegraphics[width=\columnwidth]{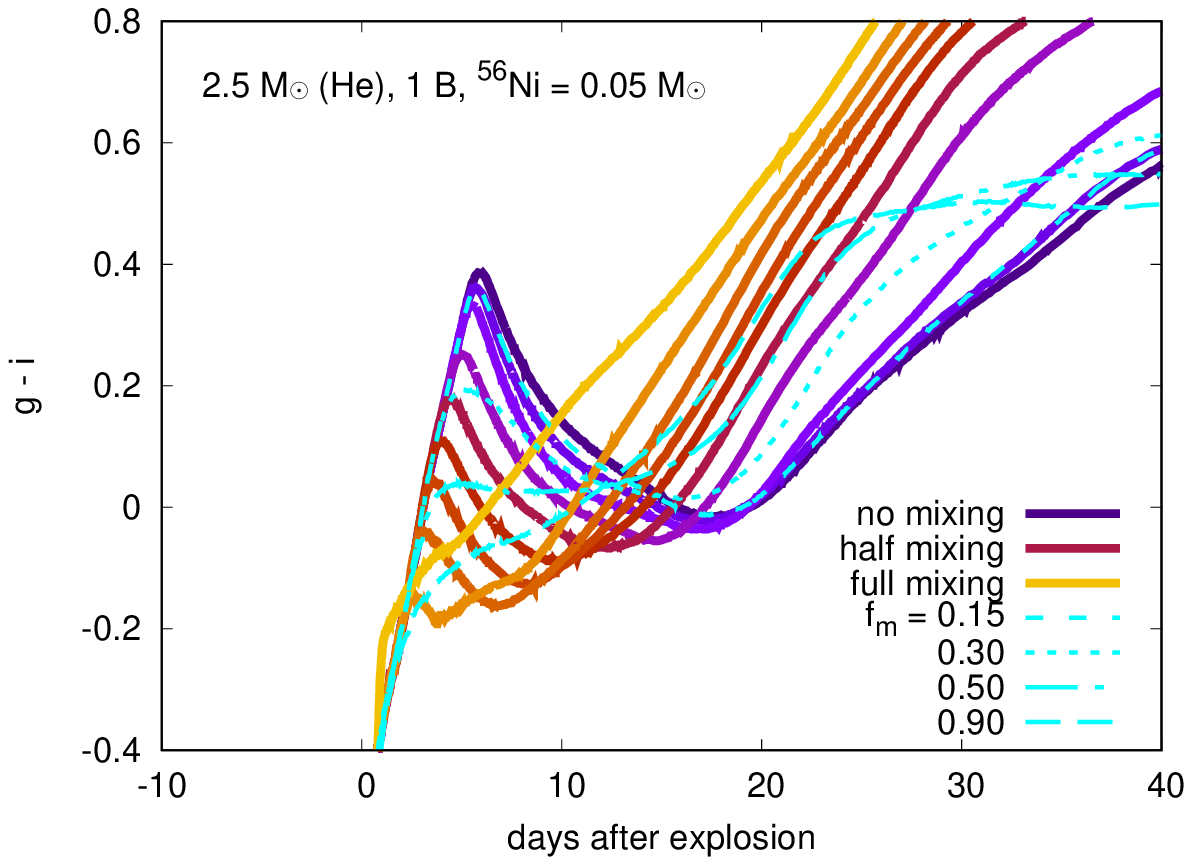}
	\includegraphics[width=\columnwidth]{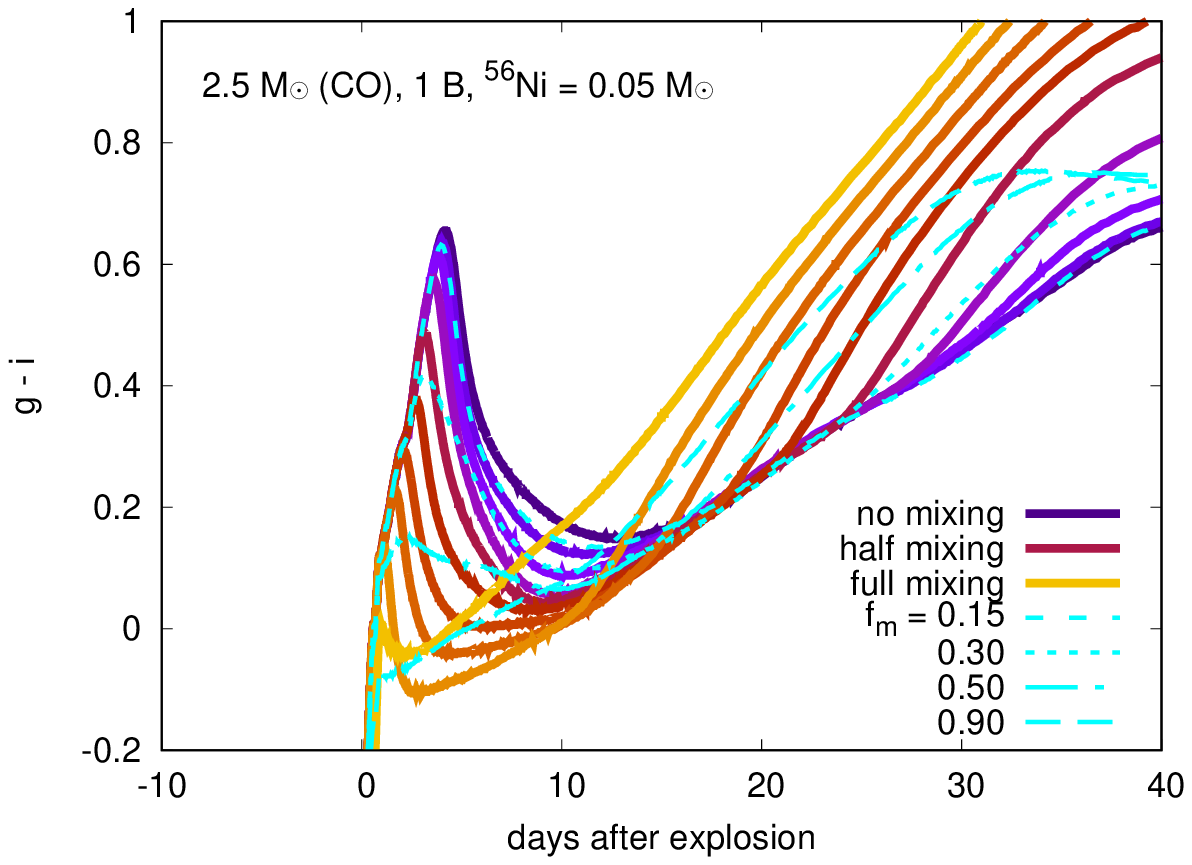}
    \caption{
    Color ($g-i$) evolution of the models presented in Fig.~\ref{fig:m2p5_1b_lc}.
    }
   \label{fig:m2p5_1b_gmi}
\end{figure}

\subsection{Color}\label{sec:color}
Fig.~\ref{fig:m2p5he_1b_gi} shows the \textit{g} and \textit{i} band LCs of our helium star explosion models. As previously pointed out by \citet{yoon2019mixing}, the effect of mixing strongly appears in the bluer bands during the recombination phase. This is due to the difference in the evolution of the photospheric temperature caused by the timing of the \Ni\ heating (Fig.~\ref{fig:m2p5_1b_temp}).
The peak magnitudes depends on the degree of \Ni\ mixing because of the difference in the temperature evolution.

As pointed by \citet{yoon2019mixing}, the effect of the difference in the photospheric temperature evolution caused by the degree of \Ni\ mixing can be easily identified in the color evolution observationally. Fig.~\ref{fig:m2p5_1b_gmi} presents the $g-i$ color evolution. The less mixed models become significantly redder because of the delay of the \Ni\ heating in both helium and carbon+oxygen star explosion models.

\begin{figure}
	\includegraphics[width=\columnwidth]{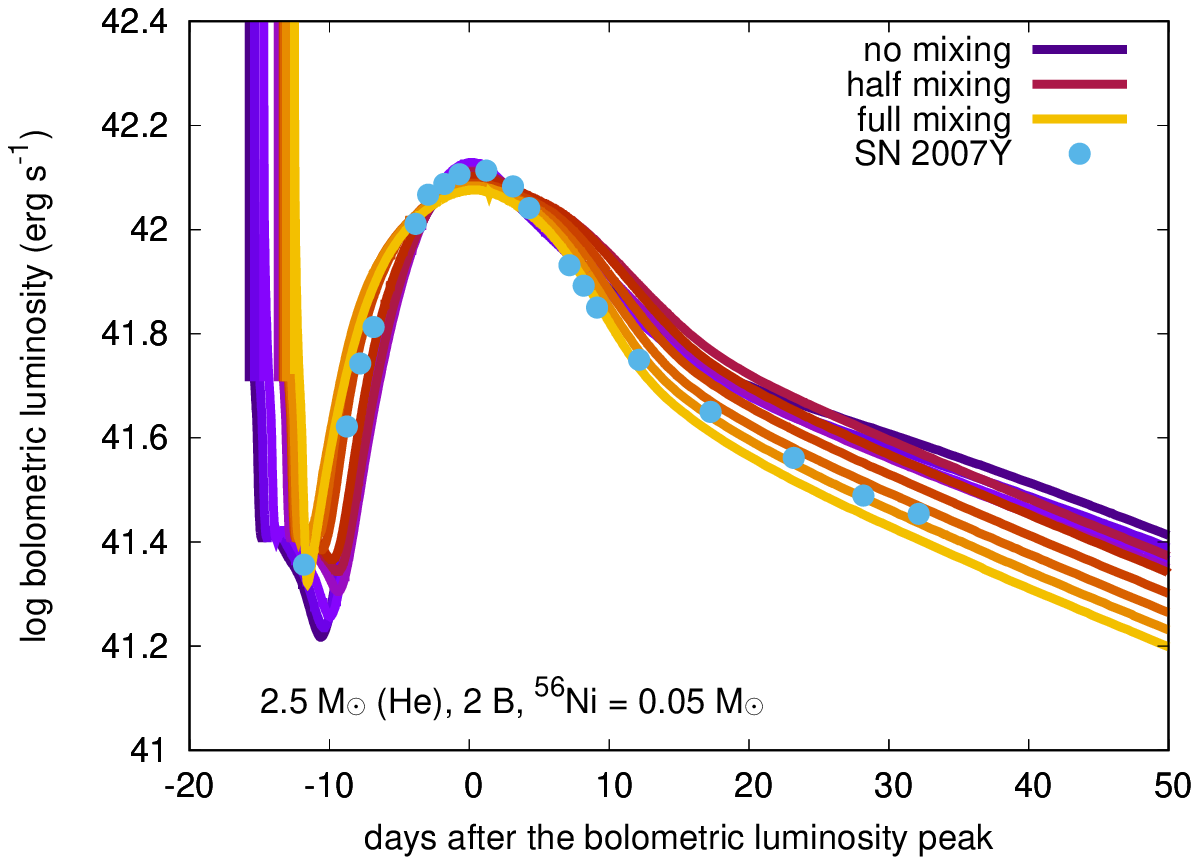}
	\includegraphics[width=\columnwidth]{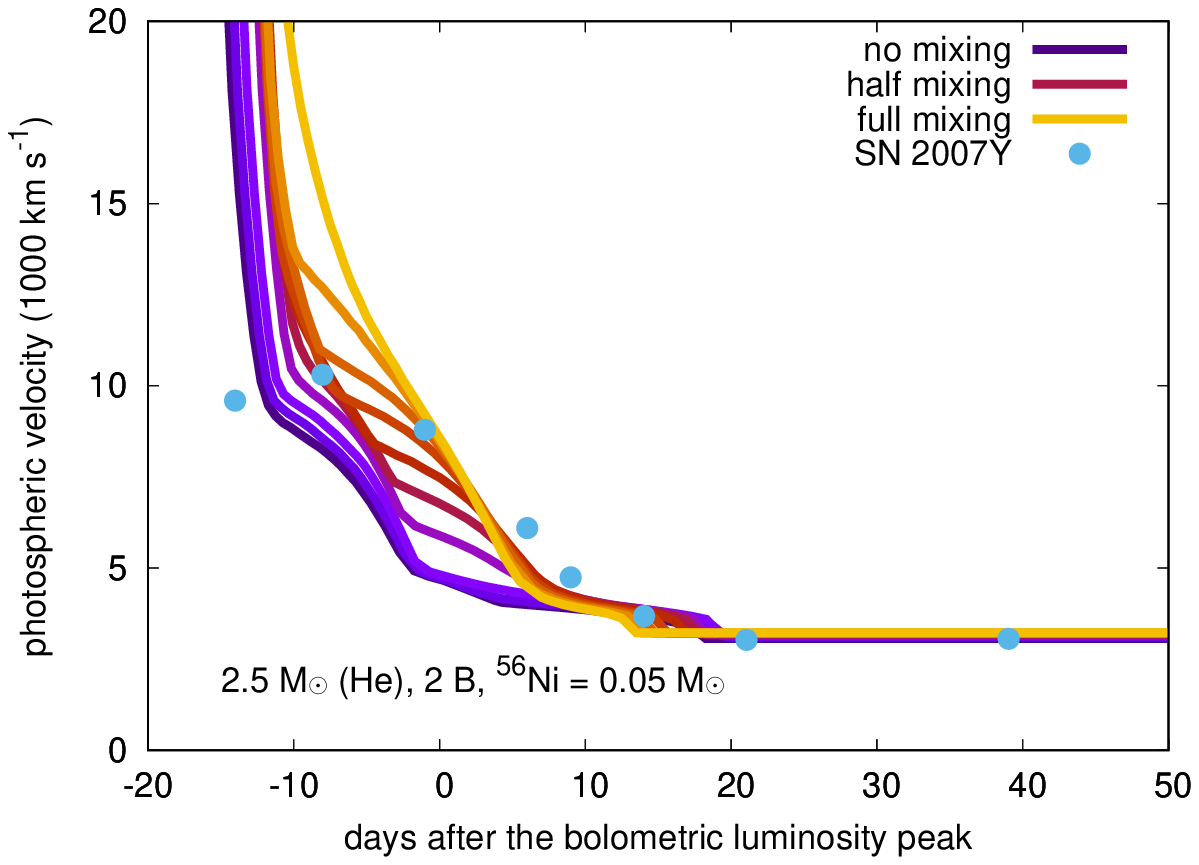}
	\includegraphics[width=\columnwidth]{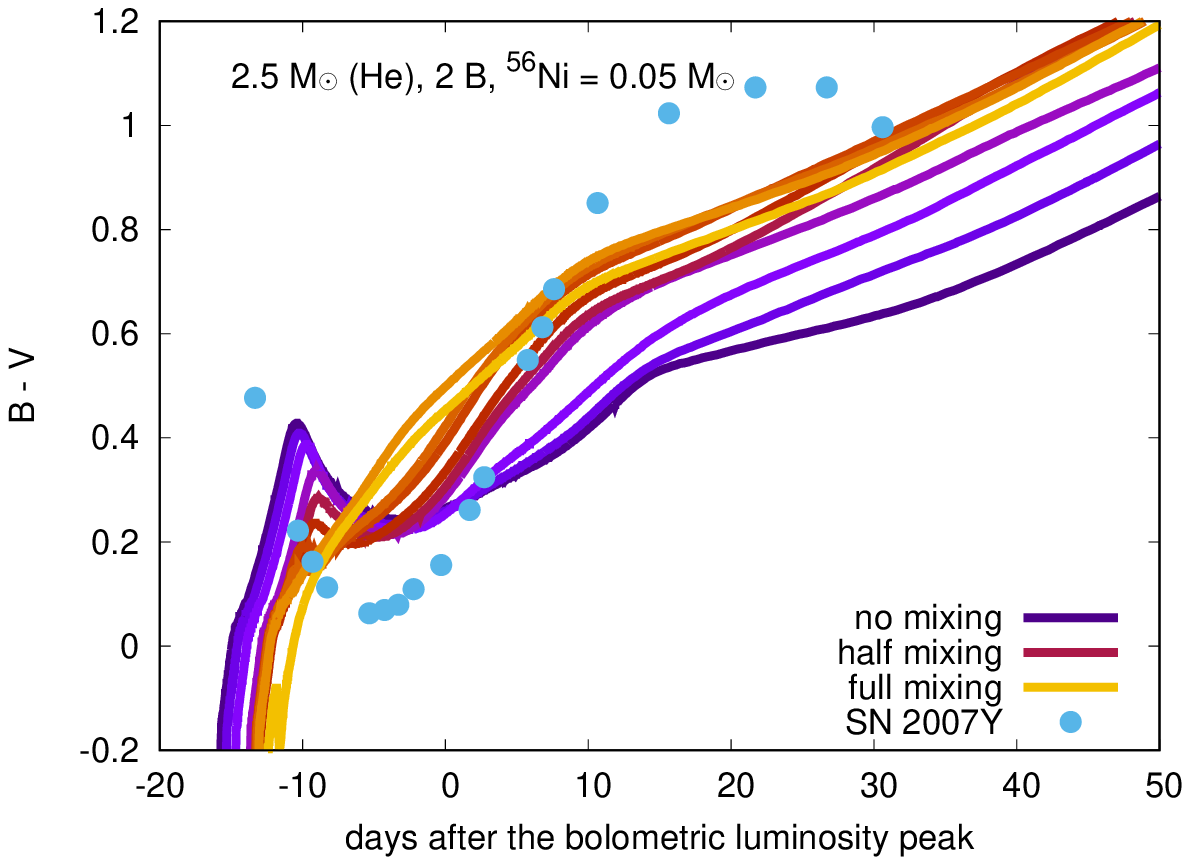}	
    \caption{
    Bolometric LC (top), photospheric velocity (middle), and color (bottom) evolution of SN~2007Y \citep{stritzinger2009sn2007y} and their comparison with the synthetic explosion models from the helium star progenitor having the ejecta mass of 2.5~\Msun, explosion energy of 2~B, and the \Ni\ mass of 0.05~\Msun. The photospheric velocity of SN~2007Y is estimated by the Fe~\textsc{ii} $\lambda 5169$ line.
    }
   \label{fig:sn2007y}
\end{figure}

\section{Discussion and conclusions}\label{sec:discussion}
We have shown that the early photospheric velocity evolution in stripped-envelope SNe is strongly affected by the \Ni\ mixing in the ejecta. Especially, we expect to find flattening in the photospheric velocity evolution when \Ni\ is not fully mixed in the ejecta. The duration of the flattening is related to the duration of the recombination phase. The less mixed models have the longer recombination phase and therefore have the longer flattening.

Observationally, the velocity of Fe~\textsc{ii} $\lambda 5169$ is used as a tracer of photospheric velocity \citep{branch2002}. Fe~\textsc{ii} $\lambda 5169$ velocity evolution in stripped-envelope SNe has been summarized in, e.g., \citet{liu2016stripped,modjaz2016stripped,taddia2018cspstripped}. To trace the duration of the flattening which differs significantly depending on the degree of mixing, the photospheric velocity information more than 10~days before the LC peak is required (Section~\ref{sec:results}). Spectroscopic observations for stripped-envelope SNe at such an early time are still very rare and are encouraged to investigate the \Ni\ mixing in stripped-envelope SNe.

Still, there are several stripped-envelope SNe with the early photospheric velocity information. For example, two SNe~Ic, i.e., SN~1983V \citep{clocchiatti1997sn1983v} and PTF12gzk \citep{ben-ami2012ptf12gzk}, in the samples summarized by \citet{modjaz2016stripped} have the Fe~\textsc{ii} velocity information before 10~days before the $V$ band LC peak and they both show the velocity increase in the earliest phases. The velocity increase in carbon+oxygen star explosions is found only in the models with relatively small degrees of mixing in our models (Fig.~\ref{fig:m2p5_1b_vel}). SNe~Ib with the early Fe~\textsc{ii} information do not have as clear flattening as in the two SNe~Ic \citep{taddia2018cspstripped,liu2016stripped}. However, one SN~Ib, SN~2007Y, does show an increase of the Fe~\textsc{ii} velocity at the earliest phases \citep{stritzinger2009sn2007y}. Using the helium star progenitor models we have, we search for the SN models providing a reasonable fit to SN~2007Y. Fig.~\ref{fig:sn2007y} shows the comparison between our models and SN~2007Y. We find that our helium star explosion models with $\Mej = 2.5~\Msun$, $\Eej=2~\mathrm{B}$, and the \Ni\ mass of 0.05~\Msun\ in Fig.~\ref{fig:m2p5he_?b_vel_peak} provide reasonable fits to the bolometric LC and photospheric velocity as shown in Fig.~\ref{fig:sn2007y}. The models with moderate degrees (roughly half) of \Ni\ mixing explain the photospheric velocity evolution well. Although the first photospheric velocity point is not well reproduced, the further velocity flattening would only be caused by less degree of mixing and our conclusion that the ejecta of SN~2007Y is mixed moderately would not be altered.
Our models do not fully reproduce the color evolution, but the ``U''-shaped color evolution found in SN~2007Y is consistent with a relatively small degree of \Ni\ \citep{yoon2019mixing}.
Obtaining a perfect fit to SN~2007Y is beyond the scope of this paper. 

Constraining the degree of \Ni\ mixing in the stripped-envelope SN ejecta is an important step towards identifying the physical mechanism responsible for the mixing, which is closely linked with the nature of the explosions. The major suggested mechanism to initiate the mixing in SN ejecta is the Rayleigh-Taylor mixing \citep[e.g.,][]{wongwathanarat2017,utrobin2019}, but having \Ni\ mixing in the entire SN ejecta by the Rayleigh-Taylor mixing is challenging \citep[e.g.,][]{basko1994,tanaka2017}. Also, we do not expect a strong degree of mixing by the Rayleigh-Taylor mixing in helium-free explosion. By constraining the degree of \Ni\ mixing through the early-phase photometric and spectroscopic information in stripped-envelope SNe, we can constrain the mixing mechanism in stripped-envelope SNe that are closely linked to their explosion mechanism.

\section*{Acknowledgements}
This work is supported by the Grants-in-Aid for Scientific Research of the Japan Society for the Promotion of Science (JP17H02864, JP17H01130, JP17H06364, JP18K13585, JP18H01212, JP19K14770, JP20H00174).
S.B. is supported by grant RSF 19-12-00229 in his work on the
supernova simulations with STELLA code.
Numerical computations were in part carried out on PC cluster at Center for Computational Astrophysics (CfCA), National Astronomical Observatory of Japan.

\section*{Data availability}
The data underlying this article will be shared on reasonable request to the corresponding author.




\bibliographystyle{mnras}
\bibliography{mnras} 







\bsp	
\label{lastpage}
\end{document}